\documentclass[useAMS,usenatbib,fleqn]{mn2e}

\usepackage{times}
\usepackage{graphicx}
\usepackage{epsfig}
\usepackage{amssymb,amsmath}
\usepackage{aas_macros}

\title[Transient jets from magnetic reconnection]{Transient jet formation and state transitions from large-scale magnetic reconnection in black hole accretion discs}
\author[Dexter et al.]{Jason Dexter$^{1}$\thanks{E-mail: 
jdexter@berkeley.edu}, Jonathan C. McKinney$^{2}$, Sera Markoff$^{3}$, and Alexander Tchekhovskoy$^{4}$\\
$^{1}$Departments of Physics and Astronomy, University of California, Berkeley, CA 94720-3411, USA\\
$^{2}$Physics Department and Joint Space Science Institute, University of Maryland, College Park, MD 20742, USA\\
$^{3}$Astronomical Institute ``Anton Pannekoek'', University of Amsterdam, Postbus 94249, 1090 GE Amsterdam, The Netherlands\\
$^{4}$Lawrence Berkeley National Laboratory, 1 Cyclotron Rd, Berkeley, CA 94720, USA; Einstein Fellow
}
\begin{document}
\pagerange{\pageref{firstpage}--\pageref{lastpage}} \pubyear{2013}
\maketitle

\label{firstpage}

\begin{abstract}
Magnetically arrested accretion discs (MADs), where the magnetic pressure in the inner disc is dynamically important, provide an alternative mechanism for regulating accretion to what is commonly assumed in black hole systems. We show that a global magnetic field inversion in the MAD state can destroy the jet, significantly increase the accretion rate, and move the effective inner disc edge in to the marginally stable orbit. Reconnection of the MAD field in the inner radii launches a new type of transient outflow containing hot plasma generated by magnetic dissipation. This transient outflow can be as powerful as the steady magnetically-dominated Blandford-Znajek jet in the MAD state. The field inversion qualitatively describes many of the observational features associated with the high luminosity hard to soft state transition in black hole X-ray binaries: the jet line, the transient ballistic jet, and the drop in rms variability. These results demonstrate that the magnetic field configuration can influence the accretion state directly, and hence the magnetic field structure is an important second parameter in explaining observations of accreting black holes across the mass and luminosity scales.
\end{abstract}


\section{Introduction}

A black hole accreting magnetic field with a consistent sign of magnetic flux reaches a limit where the magnetic pressure at the black hole resists continued accretion \citep{bisnovatyikoganruzmaikin1974,bisnovatyikoganruzmaikin1976,igumenshchevetal2003,narayanetal2003}. The accretion process in this limit is mediated by instabilities in the black hole magnetosphere, and the magnetorotational instability \citep{mri}, typically thought to cause angular momentum transport in black hole accretion discs, is marginally suppressed \citep[][MTB12]{mckinneyetal2012}. 

\citet{narayanetal2003} predicted that such a ``magnetically arrested'' disc (MAD) could be an extremely efficient engine. This was recently confirmed in general relativistic MHD simulations  \citep[][MTB12]{tchekhovskoyetal2011}, where the \citet[][BZ]{blandfordznajek1977} jet efficiency (energy expelled vs. accreted by the black hole), can be $\gtrsim 150-250\%$ at high black hole spin. Reaching this limit requires only modest coherent vertical magnetic field, suggesting that it could be generic to galactic nuclei and binary systems (MTB12). 

\citet{igumenshchev2009} argued that a polarity inversion in the accreted magnetic field in the MAD state could trigger the observed state transitions in black hole X-ray binaries (BHBs). This work was based on 2D MHD simulations, which cannot accurately capture the MAD state due to the absence of the non-axisymmetric modes which dominate the accretion in this state. In order to more accurately study the physical outcome of such an inversion, MTB12 set up numerical experiments in several of their 3D, high resolution, general relativistic simulations of MAD accretion flows. Their initial conditions contained large scale field inversions, where adjacent magnetic field loops in the initial density distribution have opposite magnetic polarity.  The magnetic flux in each loop was chosen to be much more than required to establish the MAD state. As accretion proceeded, a MAD state was established, reached quasi-steady state, and then a large amount of coherent field of the opposite polarity was accreted. MTB12 studied the quasi-steady structure of the MAD state in these simulations, but did not analyze the outcome of these polarity inversions.

In this Letter, we study the evolution of the accretion flow and jet during a large-scale field inversion experiment carried out by MTB12. We demonstrate that the accreted magnetic field configuration can indeed change important properties of the accretion flow including the mass accretion rate, the disc geometry, and the effective inner disc edge (\S\ref{sec:mads}). We further show that the reconnection of the MAD magnetic field destroys the BZ jet and launches a new type of transient outflow. We discuss possible observational implications of these results, particularly for BHBs and their state transitions, in \S \ref{sec:observ-impl}.

\begin{figure}
\begin{center}
\includegraphics[scale=0.75]{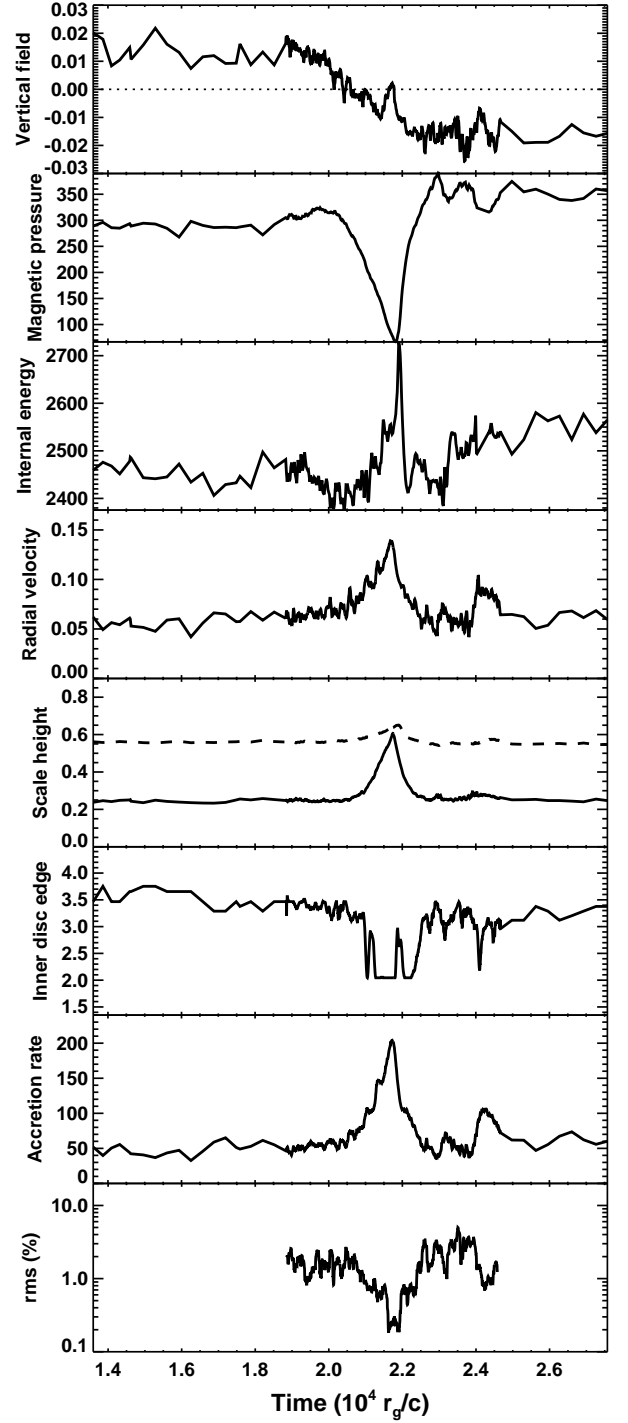}
\caption{\label{shellavg}Time evolution of several quantities during the A0.94BfN40 simulation from MTB12. The inner disc edge is in units of $r_g$, the scale height is dimensionless, and the radial velocity is in units of $c$. All other quantities use (arbitrary) code units. The pressure, radial velocity, and vertical field strength are measured at $r=5\hspace{2pt} r_g$. The accretion rate is measured at $r=3\hspace{2pt} r_g$. The scale height is measured at $r = 3 r_g$ (solid line) and $5 r_g$ (dashed line). The magnetic and internal energies are measured at  $r=25\hspace{2pt} r_g$.  During the inversion, the vertical field polarity flips in sign. The magnetic pressure drops sharply as the two field polarities mix and reconnect, resulting in an increase in thermal energy. The accretion rate increases due to an increase in the radial velocity, the inner disc is no longer compressed by pressure at the disc-jet interface, the inner disc edge moves towards the black hole, and the rms variability drops. These effects are reversed as the MAD state is re-established.}
\end{center}
\end{figure}

\section{A field inversion in a magnetically arrested black hole accretion disc}
\label{sec:mads}

The fiducial simulation presented in MTB12 (named A0.94BfN40 in their Table 3) is the highest resolution, longest duration 3D simulation of a MAD configuration to date, providing a wealth of information about the physics of magnetized accretion. We extend their analysis by studying the evolution of this simulation to late times, following a polarity inversion in the accreted magnetic field. We use area-integrated and shell-averaged radial profiles in Boyer-Lindquist coordinates to study the evolution of various quantities. The density-weighted shell-average of a quantity Q is defined as:

\begin{equation}
\langle Q \rangle = \frac{\int dA_{\theta \phi} \rho Q}{\int dA_{\theta \phi} \rho},
\end{equation}

\noindent where $\rho$ is the density, $dA_{\theta \phi} = d\theta d\phi \sqrt{-g}$, and $g$ is the metric determinant. 

Long after a quasi-steady MAD state is established in the inner disc ($r < 25 r_g$, where $r_g = G M / c^2$ is the gravitational radius and $M$ is the black hole mass), material is accreted with the opposite magnetic field polarity as that on the black hole. Figure \ref{shellavg} shows density-weighted shell-averaged radial profiles of various quantities before, during, and after the ensuing field inversion at $t \sim 20000 \hspace{2pt} r_g / c$. The quantities considered include the vertical magnetic field strength, taken as the azimuthally-averaged $b^\theta (\theta=\pi/2)$, where $b^\mu$ is the magnetic field four-vector measured in Heaviside-Lorentz units; the total pressure: $p_g + b^2 / 2$, where $p_g$ is the gas pressure; and the radial velocity, $v^r \equiv u^r / u^t$. As in MTB12, the disc scale height, $\theta_d$, is defined as:

\begin{eqnarray}
\theta_d = \langle \left ( \theta - \theta_0 \right)^2 \rangle^{1/2},\hspace{12pt}
\theta_0 = \pi/2 + \langle (\theta - \pi/2)\rangle,
\end{eqnarray}

\noindent where $\theta_0$ is the midplane location. The measured rms variability is calculated as the difference between the actual accretion rate curve and a smoothed version in order to remove secular trends on longer timescales. It is only calculated during the field inversion. This procedure enables accurate estimates of fluctuations during the inversion, but leads to small rms values at all times (e.g., rms $\simeq 1\%$ compared to $\simeq 30\%$ in the raw accretion rate curve in the MAD state).

The mass and energy fluxes are,

\begin{eqnarray}
\dot{M} = \int dA_{\theta \phi} \rho u^r, \hspace{8pt} \dot{E} = \int dA_{\theta \phi} T^r_t, 
\end{eqnarray}

\noindent where $T^\mu_\nu$ is the stress energy tensor. We will further use the fluxes of electromagnetic, $T^{r, \rm EM}_{t} = b^2 u^r u_t - b^r b_t$,  kinetic, $T^{r, \rm KE}_t = \rho u^r (1 + u_t)$, and thermal, $T^{r, \rm EN}_t = (u_g + p) u^r u_t$ energy where $u_g$ is the internal energy density. 

The inner disc radius is defined as the minimum of the stress, infall, and turbulent velocity measures from \citet{krolikhawley2002}, but is not allowed to move inside the ISCO. The jet efficiency is defined as,

\begin{equation}\label{eq:2}
\eta_j \equiv \frac{\langle \dot{M} \rangle - \langle \dot{E}_j \rangle}{\langle \dot{M} \rangle},
\end{equation}

\noindent where the subscript $j$ denotes that the jet power includes only the energy fluxes in the jet and wind (with the jet being the dominant contribution). In the MAD state, the jet can be robustly defined as magnetically-dominated regions with $b^2 / \rho c^2 > 1$ and the wind as regions with $b^2 / \rho c^2 < 1$ and $2 p_g / b^2 < 1$ (MTB12). However, during the field inversion these definitions do not capture the region of interest. We instead define the jet and wind as regions near the pole with $\theta < 10^\circ$ and $30^\circ$, and measure jet powers at $r=50 r_g$ where both choices for the wind and jet region give consistent results.

\subsection{Quasi-steady MAD state}

Prior to the field inversion event ($t \lesssim 19000 r_g / c$), the accretion flow is in the quasi-steady MAD state analysed in detail by MTB12. A strong, coherent vertical field is present in the inner disc. The jet is powerful, and clearly extracting black hole spin energy (efficiency $\simeq 250\%$). The scale height at $r = 3 r_g$ is significantly smaller ($h/r \simeq 0.25$) than that at $r=5 r_g$ ($h/r \simeq 0.6$) due to compression by magnetic pressure from the surrounding jet magnetosphere. The MRI is marginally suppressed in this state, and accretion proceeds through instabilities at the jet-disc interface, leading in this simulation to fluctuations in the mass accretion rate and quasi-periodic oscillations in other dynamical quantities (e.g., the jet power). For the thinner discs ($\theta_d \simeq 0.3$) studied in \citet{tchekhovskoyetal2011}, the rms noise is larger, and no QPOs are present.

\subsection{Field inversion}
The evolution during the magnetic polarity inversion ($t \gtrsim 19000 r_g / c$) is studied here for the first time. As the opposite polarity loop is accreted, the two field polarities mix and reconnect throughout the disc. This can be seen in the top two panels of Figure \ref{shellavg}: the vertical field in the inner disc changes sign as the inversion proceeds, while the internal energy increases as the magnetic pressure decreases.

Subsequently, the inner disc expands vertically due to the removal of magnetic pressure from the jet (e.g., the scale height at all radii expands to that imposed by the initial conditions, $\theta_d \simeq 0.6$), and therefore the radial velocity increases. The accretion rate then increases in response to the larger radial velocity. The effective inner disc edge moves closer to the black hole as the disc more closely resembles a standard MRI state. In the (short lived for these initial conditions, $\Delta t \approx 2000 r_g / c$) MRI state, the QPOs disappear and the rms variability in all dynamical quantities drops (bottom panel of Figure \ref{shellavg}).

These results demonstrate that the magnetic field geometry accreted can greatly affect the evolution of the disc, and even control the accretion rate, which is usually considered to be an independent parameter. The reason for the accretion rate change is that the radial velocity allowed by magnetic Rayleigh-Taylor and interchange instabilities \citep{stonegardiner2007} in the MAD state is different from that set by the MRI alone. Once the field inversion destroys the built up magnetically-dominated jet, the MRI determines the radial velocity and in general gives a different quasi-steady accretion rate. The MAD accretion rate should always be smaller than that in the MRI state, since the strong magnetosphere at the black hole in the MAD state is actively impeding accretion.

\begin{figure}
\includegraphics[scale=0.65]{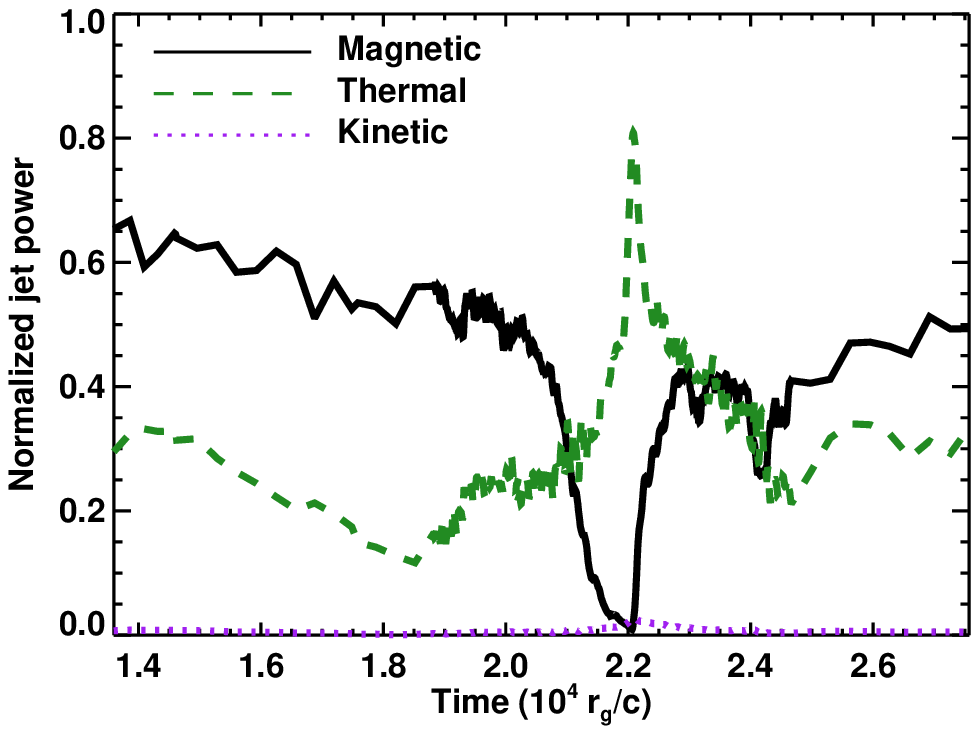}\\
\includegraphics[scale=0.65]{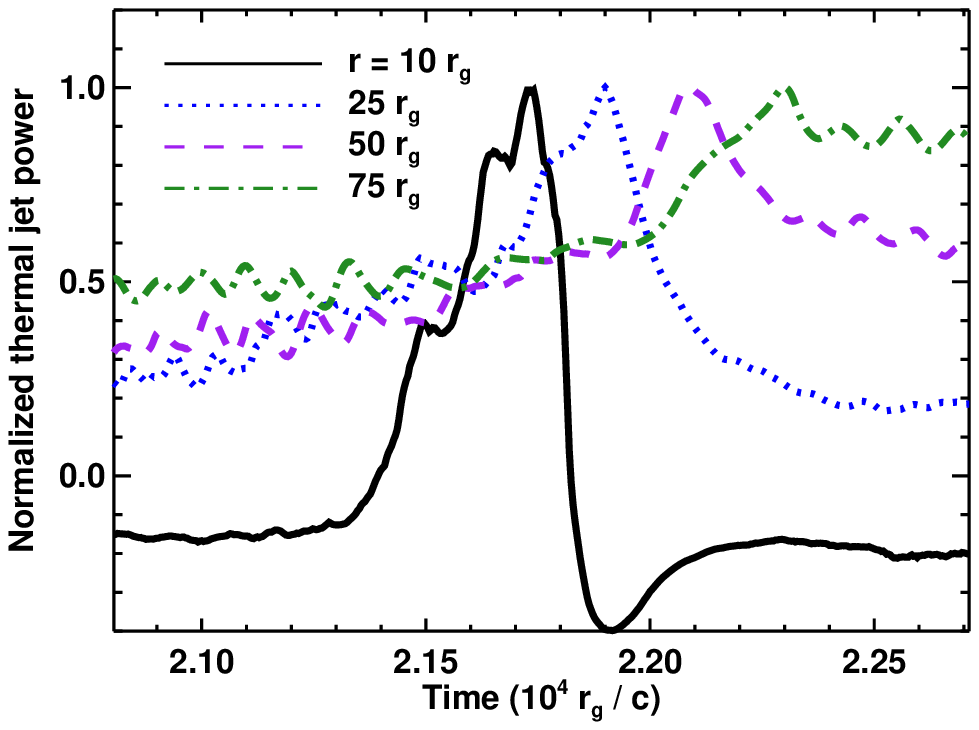}
\caption{\label{jetenergy}Magnetic, thermal, and kinetic energy flux in the polar region ($\theta < 30^\circ$) at $r = 50 r_g$ (top) and thermal energy flux measured at several radii vs. time (bottom) during a field inversion in a magnetically arrested accretion disc. During the field inversion, the magnetic energy is converted into thermal and kinetic energy flux. This transient jet propagates outwards (velocity $\simeq 0.1 c$) with comparable power and velocity at small radius to the steady magnetically-dominated BZ jet. The negative energy flux at small radius in the bottom panel indicates inflow rather than outflow.}
\end{figure}

\subsection{Transient jet formation}

The production of magnetically-dominated jets in GRMHD simulations requires coherent vertical field threading the black hole \citep{beckwithetal2008jet,mckinneyblandford2009}. As field of the opposite polarity is accreted, the vertical field on the black hole decreases due to magnetic reconnection. This in turn destroys the magnetically-dominated MAD state jet. Figure \ref{jetenergy} shows the magnetic, kinetic, and thermal energy fluxes in the jet as functions of simulation time. The decrease of the Poynting flux corresponds to the destruction of the BZ jet.

Following the inversion, the kinetic and thermal fluxes suddenly increase. These increases correspond to a transient outflow, which propagates outwards as shown in the bottom panel of Figure \ref{jetenergy}. The implied speed from the delay between the peaks in flux at different radii is $\simeq 0.1 c$, about twice the velocity of the MAD jet at $r = 50 r_g$. These measured velocities are sub-relativistic because the jet is still accelerating at these radii. The terminal Lorentz factor gives an upper limit to the velocity at large radius, outside the simulation domain \citep{mckinney2006}: $\Gamma_{\infty} = -\dot{E}_j/\dot{M}_j$. The upper limit to the Lorentz factor of the transient outflow, $\Gamma_{\infty} \simeq 1.3-2.0$, is smaller than that of the BZ jet: $\Gamma_{\infty} \simeq 3-30$.  Further study will be required to determine whether these transient outflows can become ultrarelativistic at large radii.

The most natural energy source for this outflow is magnetic energy converted into thermal energy during the field inversion. Most of this energy in the simulation is contained within $r \lesssim 10 r_g$, but the vast majority is lost to the black hole. Between $r = 10-50 r_g$, the thermal energy increases by roughly the same amount as the magnetic energy drops during the conversion, implying efficient heating of the gas from reconnection. Therefore, the outflow is powered by reconnection in the disc rather than at the black hole. The energy in the outflow also increases as a function of radius, indicating that a wide range of radii contribute to its power. 

The transient outflow velocity is $\simeq 0.05-0.1 c$ over the radial range ($r \lesssim 200 r_g$) where it can be followed. It is not clear what sets the velocity. In the simulation we study, the outflow is accelerated by a BZ jet which re-forms following the inversion. However, the material is already unbound and flowing outwards before the BZ jet re-forms. The outflow duration at $r = 50 r_g$ is shorter than expected from reconnection liberating thermal energy on the inflow time, and the velocity is smaller than the local escape speed. It is possible that faster material from smaller radii sweeps up that at larger radius, but the velocity is roughly constant with radius. 

Unlike the BZ jet ($P \sim a^2$ at low spin), the transient jet power does not appear to depend on black hole spin. The $a=0$ MTB12 simulation with the same initial condition has no BZ jet but forms a transient outflow with comparable speed and power. Therefore, large-scale reconnection may produce powerful jets even at low black hole spin, which has not previously been possible in global simulations. However, the lower spin MTB12 simulations were not run long enough for the second polarity inversion to occur, which we have studied here for their fiducial simulation (which required $\sim 10^7$ cpu-hours). This second inversion occurs after a quasi-steady MAD state is achieved out to $r > 25 r_g$, and is therefore more likely to give robust results. We plan to study the properties of inversions and transient outflows as a function of black hole spin and disc thickness.

\begin{figure}
\begin{center}
\includegraphics[scale=0.65]{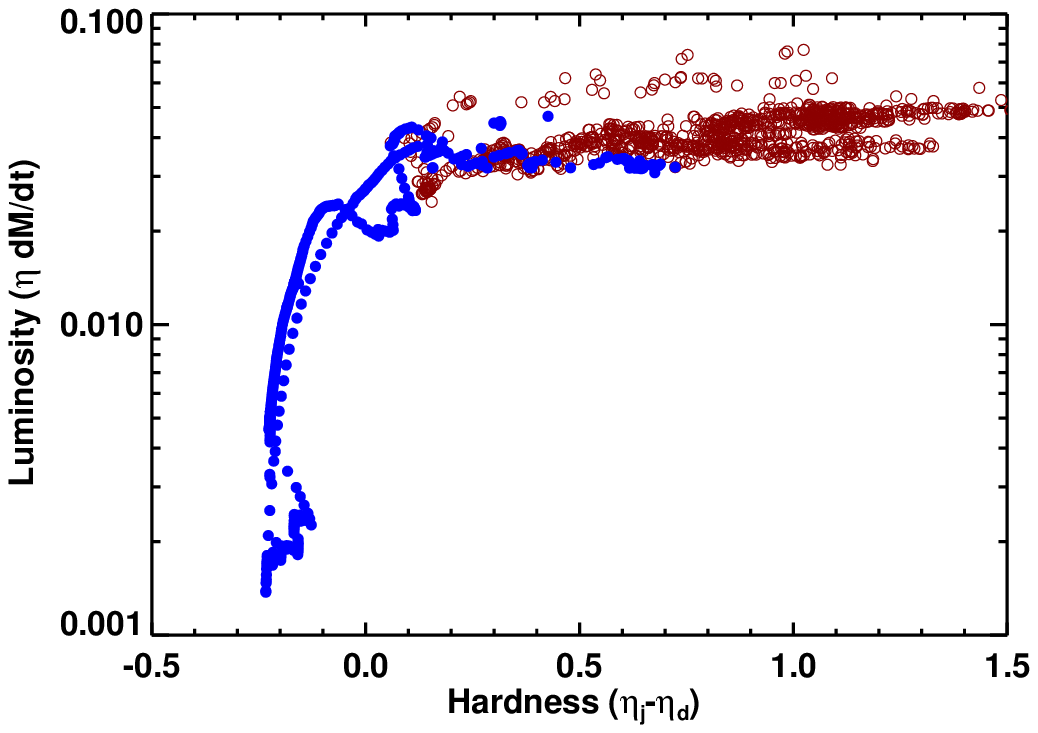}\\
\includegraphics[scale=0.65]{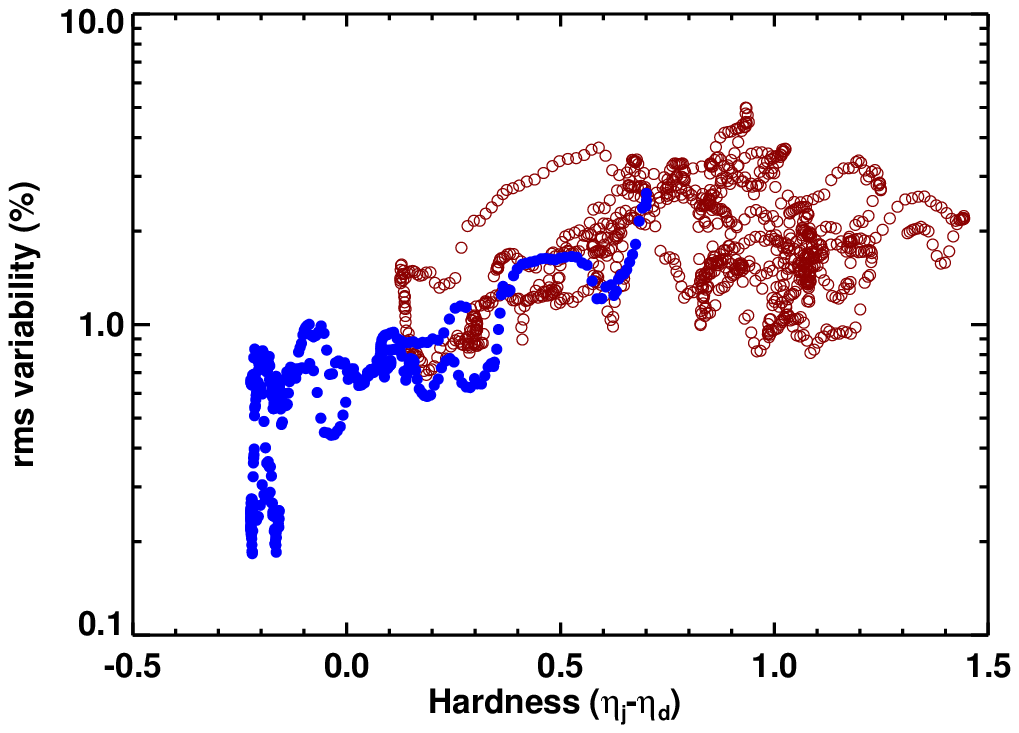}
\caption{\label{hardnessluminosity}Proxies for luminosity vs. hardness (top) and rms variability vs. hardness (bottom) sampled during (solid) and before/after (open) a field inversion in a simulation of a magnetically arrested accretion flow. The hardness is taken to be the difference between jet (equation \ref{eq:2}) and disc efficiencies, while the total luminosity is the sum of jet and disc efficiencies multiplied by the accretion rate. The rms variability values are low because they are measured by subtracting the accretion rate curve from a smoothed version, in order to remove secular variations. The field inversion causes a transition between a more variable ``hard'' state and a quieter ``soft'' state.}
\end{center}
\end{figure}

\section{Observational Implications}
\label{sec:observ-impl}

We have demonstrated that global magnetic field polarity inversions in magnetically arrested accretion discs (MADs) can directly influence the accretion state. In the simulation studied here (the fiducial simulation first presented by MTB12), magnetic reconnection following a global field inversion destroys the steady jet. In addition, the inflow is no longer inhibited by the strong magnetic field, so that the accretion rate rises by a factor of several. The accreted magnetic field geometry, then, can control the accretion rate onto the black hole. Reconnection from the accretion of opposite polarity field converts magnetic energy in the disc into kinetic and thermal energy fluxes (Figure \ref{jetenergy}). This energy flux propagates outwards, at $\simeq 0.1c$ out to $100 r_g$ in the simulation studied here, with a mildly relativistic terminal Lorentz factor of $1.3$. This is a new type of relativistic outflow from a black hole accretion disc, whose power is comparable to that of the magnetically-dominated Blandford-Znajek jet in this simulation. This method for dissipating jet Poynting flux should operate in MRI discs as well as the MAD disc studied here.

Our results may have several implications for observed accreting black holes. At least two types of jets are observed in BHBs: a steady, compact jet \citep{mirabeletal1992,fender2001} is seen in the hard spectral state  \citep[e.g.,][]{remillardmcclintock2006}, while a transient, ballistic jet is seen during transitions from the hard to soft state  \citep{mirabelrodriguez1994}. Neither type of jet is observed in the soft state \citep{fenderetal1999}. The transient jet seen here is also produced during a change in state of the accretion disc, from the MAD to the MRI state. It is therefore tempting to associate the hard and soft states with MAD and MRI accretion flows, as first suggested by \citet{igumenshchev2009}. The hysteresis seen in black hole outbursts would then correspond to whether or not the accretion flow is magnetically arrested.

Several of our results support this scenario. The MAD state is associated with a steady jet, while MRI jets are either absent or weak, depending on the spin and disc thickness \citep{pennaetal2010}. The MAD jet power is proportional to the accretion rate \citep{tchekhovskoyetal2011}, naturally explaining the observed radio/X-ray correlations in BHBs whether the hard X-rays are produced in the disc \citep[e.g.,][]{magdziarzzdziarski1995,esinetal1997} or the jet  \citep[e.g.,][]{markoffetal2001,markoffetal2005}. The recently discovered radio loud and quiet tracks in the hard state \citep[e.g.,][]{coriatetal2011} could correspond to weak (strong) jets from an MRI (MAD) accretion state. The large MAD jet efficiences could explain the high X-ray luminosities reached in the hard state, which are difficult to explain using radiatively inefficient accretion.

The transition from the MAD to the MRI state also qualitatively resembles the observed hard to soft state transitions in BHBs. As the steady MAD jet is destroyed during the field inversion, the rms variability in many quantities (e.g., the accretion rate) drops sharply, and a new type of transient outflow is launched, in qualitative agreement with observations. To demonstrate this idea, we show proxies for the X-ray color (hardness ratio), luminosity, and rms variability during the state transition in Figure \ref{hardnessluminosity}. Using these proxies, the inversion event shows a transition from a more variable hard state to a quieter soft state at similar luminosity, qualitatively similar to observed hardness-intensity and hardness-rms diagrams from black hole outbursts. This transition is driven by large-scale magnetic reconnection, in this simulation following a field polarity inversion. We note that reconnection could alternatively be a side effect of a different mechanism driving BHB outbursts, e.g., cooling instabilities \citep{esinetal1997,dassharma2013}, rather than the root cause. In this case it could explain the spectral state transition and the associated transient jet ejections.

If the magnetic field configuration does play an important role in BHB state transitions, then different outburst cycles in persistent and transient sources may be related to the efficiency of magnetic field transport in accretion discs. In transient sources where the soft state is reached, the disc may collapse and become geometrically thin, preventing further efficient transport of magnetic flux \citep{lubowetal1994}. In persistent sources which never reach the soft state, the MAD state could be re-established on a relatively short timescale, triggering a return to the hard state. This could explain the repeated jet ejection cycles in a source like GRS 1915+105 \citep[e.g.,][]{neilsenlee2009} or 3C 111 \citep{chatterjeeetal2011}. These could also be the result of partial reconnection events, which power a transient outflow but do not completely destroy the MAD state.

Our estimated transient jet power depends on the magnetic energy density and the timescale over which it is dissipated. It does not seem to depend on black hole spin, and therefore naively we predict that the observed transient jet power should be roughly independent of black hole spin, in agreement with some analyses \citep{russelletal2013} and potentially at odds with others \citep{narayanmcclintock2012,steineretal2013}. This issue is complicated by the fact that the observed radio emission comes from much larger scales than the jets studied in the simulation. The propagation of these new transient outflows to larger scales, including their dynamical and radiative properties, should be studied in future work. 

The radiative properties of the new transient jets described here are particularly interesting. Simulated BZ jets contain an ``empty funnel'' \citep{devilliersetal2005,mckinney2006}. In order to calculate observables from BZ jets, it is necessary to invoke some source of particles, either from a physical process \citep[e.g., pair-production,][]{moscibrodzkaetal2011} or as an ad-hoc prescription \citep{broderickmckinney2010,dexteretal2012,moscibrodzkafalcke2013}. In contrast, the transient jets are dominated by particle energy generated by magnetic reconnection.

MAD accretion flows, jets, and transient outflows may play a role in a variety of BH systems. \citet{sikorabegelman2013} suggested that the radio loud/quiet dichotomy in active galactic nuclei could be due to the presence or absence of sufficient coherent magnetic field to develop a MAD accretion state. This is similar to our association of MAD and MRI accretion with the hard and soft states of BHBs. Alternatively, the dichotomy could be due to the type of jet present in the system: the steady MAD jet, or a  transient outflow powered by magnetic reconnection. \citet{tchekhovskoyetal2013} showed that many observed properties of the putative tidal disruption flare Swift J1644+57 can be explained by a MAD state jet. Some of the large amplitude variability occurring in that event at early times could alternatively be explained by transient outflows triggered by magnetic reconnection. Reconnection could also produce similar variability seen in long gamma ray bursts \citep{progazhang2006,mckinneyuzdensky2012}. Our results show that field polarity inversions are one possible mechanism for triggering large-scale magnetic reconnection.

\section*{acknowledgements}
We thank R. Fender, E. Quataert, and P. Sharma for useful discussions related to this work. This work used NSF/XSEDE resources provided by NICS (Nautilus) under the award TG-PHY120005. AT was supported by NASA through the Einstein Fellowship Program, grant PF3-140115.

\footnotesize{
\bibliographystyle{mn2e}
\bibliography{master}
}
\label{lastpage}

\end{document}